\documentclass[prc,twocolumn,superscriptaddress,superscriptfootnote]{revtex4}
\usepackage{amsmath,amsfonts,amssymb,epsfig,color}

\newcommand{\SU}[2]{\ensuremath{\mathrm{SU}^{ #1 }( #2 )}}
\newcommand{\Un}[2]{\ensuremath{\mathrm{U}^{ #1 }( #2 )}}

\newcommand{\Spn}[1]{\ensuremath{\mathrm{Sp}( #1 )}}

\newcommand{\su}[2]{\ensuremath{\mathfrak{su}^{ #1 }( #2 )}}

\newcommand{\so}[1]{\ensuremath{\mathfrak{so}( #1 )}}

\newcommand{\spn}[1]{\ensuremath{\mathfrak{sp}( #1 )}}
\newcommand{\spR}[1]{\ensuremath{\mathfrak{sp}( #1, \mathbb{R} )}}

\newcommand{\half}{\ensuremath{\textstyle{\frac{1}{2}}}}

\newcommand{\sevhalves}{\ensuremath{\textstyle{\frac{7}{2}}}}

\newcommand{\dimN}{\ensuremath{{\mathcal N}}}
\newcommand{\IAS}{isobaric analog $0^+$ state}
\newcommand{\IASs}{isobaric analog $0^+$ states}
\newcommand{\CDB}{CD-Bonn}
\newcommand{\CDBt}{CD-Bonn+3terms}
\newcommand{\Gm}{GXPF1}
\newcommand{\HQ}{\ensuremath{H_Q^\perp(2)}}
\newcommand{\Hsp}{\ensuremath{H_{\spn{4}}}}
\newcommand{\fpg}{\ensuremath{1f_{5/2}2p_{1/2}2p_{3/2}1g_{9/2}} }

\newcommand{\flevel}{\ensuremath{1f_{7/2}} }

\begin{document}

\title{Underlying Symmetries of Realistic Interactions \\ and the \\
Nuclear Many-body Problem}
\author{K. D. Sviratcheva}
\author{J. P. Draayer}
\affiliation{Department of Physics and Astronomy, Louisiana State University,
Baton Rouge, Louisiana 70803, USA}
\author{J. P. Vary}
\affiliation{Department of Physics and Astronomy, Iowa State
University, Ames, IA
50011, USA}
\date{\today}

\begin{abstract}
The present study brings forward important information, within  the 
framework of spectral
distribution theory, about the types of forces that dominate  three 
realistic interactions,
\CDB, \CDBt~ and \Gm, in nuclei and their ability to account for 
many-particle effects such
as the formation of correlated nucleon  pairs and enhanced quadrupole 
collective modes.
Like-particle and proton-neutron isovector pairing correlations are
described microscopically by a model interaction with \Spn{4} 
dynamical symmetry, which is
extended to include an additional quadrupole-quadrupole interaction.
The analysis of the
results for the
\flevel level shows that both  \CDBt~ and \Gm~ exhibit a 
well-developed pairing character
compared to
\CDB, while the latter appears to build up more (less) rotational 
isovector $T=1$
(isoscalar $T=0$) collective features. Furthermore, the three 
realistic interactions are in
general found to correlate strongly with the
pairing$+$quadrupole model interaction,
especially for the highest possible isospin group of  states where 
the model interaction can
be used to provide a reasonable description of  the corresponding 
energy spectra.
\end{abstract}

%\pacs{21.60.Fw,21.30.Fe,21.10.Re,21.60.Cs}

\maketitle

%Figures included:
%scalarV.eps
%TscalarV.eps
%scalarSpectra.eps
%TscalarSpectraT1.eps
%TscalarCorrelCDB.eps
%TscalarCorrelCDBt.eps
%TscalarCorrelGm.eps

\section{Introduction}

A unified microscopic description of light to heavy nuclei requires a
comprehensive
understanding of the strong interaction and how it manifests itself
in the nuclear
medium.  Effective interaction theory attempts to model the essence
of this strong
interaction in terms of one, two, and sometimes higher (three or even
four)-body
interactions for the purpose of supporting microscopic shell-model
calculations that
target reproducing striking features of nuclei. Such are strong
pairing correlations
found near closed shells that yield to collective rotational motion
as one moves away
from shell closure as well as more subtle effects that must be
understood at a deeper
level to reproduce, for example, nuclear abundances as realized
through fast or slow
decay processes within and between nuclear species.  While good
progress is being made
toward understanding the strong force, especially through the  recent
work on lattice
quantum chromodynamics studies, much work remains to  be done.

Short such a comprehensive understanding of the strong interactions,
one way to  gain
insight into the principal characteristics of various microscopic
interactions is to
perform a detailed study of their matrix elements or common
quantities they generate,
such as eigenvectors or eigenvalues. For instance, in the same
environment two very
similar interactions are expected to yield similar patterns of their
matrix elements
(see, for example, \cite{KlingenbeckKHG77,HonmaOBM04,NegoitaVSNP05}),
as well to yield
large overlaps of corresponding eigenstates (e.g.,
\cite{Halse89,HoroiBZ01}) and close energy spectra (e.g.,
\cite{VaryY77,PovesSCN01}). On
the  other hand, a complementary comparison that is based on the
theory of spectral
distributions
\cite{FrenchR71,ChangFT71} that invokes overall correlations of two
interactions offers
a broader view on their global behavior and universal properties
\cite{HechtDraayer74,HalemaneKD78,Kota79,Rosensteel80,DraayerR82}.

The theory of spectral distributions of French and collaborators is
an alternative approach for studying effective interactions
\cite{ChangFT71,DraayerOP75,Potbhare77}
and continues to be a powerful concept  with recent 
applications in quantum chaos and
nuclear astrophysics including studies on nuclear
level densities, transition strength densities, and 
parity/time-reversal violation
(for example, see 
\cite{FrenchKPT88,KotaM94,TomsovicJHB00,GomezKKMR03,HoroiGZ04and03,Kota05}).
The significance of the method is related to the fact that low-order
energy moments over a
certain domain of single-particle states, such as the energy
centroid of an interaction
(its average expectation value) and the deviation from  that average,
yield valuable
information about the interaction that is of  fundamental importance
\cite{Ratcliff71,French72,Potbhare77,HalemaneKD78,DaltonBV79,CounteeDHK81,French83,DraayerR83a,SarkarKK87}
without the need for carrying out large-dimensional matrix diagonalization
and with little to no limitations due to the dimensionality of the 
vector space.
Within this general
framework, a simple and elegant global comparison of pairs of
interactions can be
carried in a unified way regardless of how the interactions are built
or of the models
that adopt them.  It is based on their correlation, which is a  measure that is
independent of the centroids of the interactions. This correlation
measure turns on a
comparison of the one- and two-body parts of the interactions,  and
in so doing probes
beyond the mean-field potential.

In addition, the group-theoretical foundation of the spectral
distribution methods
naturally establishes a propagation of information from nuclear
systems of two particles
to many-fermion nuclei \cite{ChangFT71,HechtDraayer74} and makes the
approach especially
suitable for studies of the goodness/breaking of  symmetries in
complex many-particle
systems
\cite{HechtDraayer74,HalemaneKD78,CounteeDHK81}. Such studies can
likewise help reduce
the dimensionality of a model space  to a tractable size by detecting
the optimal
subspace domain for a particular  many-body problem where microscopic
calculations
become feasible
\cite{FrenchR71,DraayerR83a,DraayerR83b}.

In this paper we employ the theory of spectral distributions to
compare three realistic
interactions, namely \CDB~ \cite{MachleidtSS96M01}, \CDBt~
\cite{PopescuSVN05} and \Gm~ \cite{HonmaOBM04} and two
pairing$+$quadrupole model
interactions \cite{SGD03,SGD04}. Such a study is important for
understanding the types
of forces that dominate a realistic interaction and its ability  to
describe correlated
and collective phenomena. Most significant are the  formation of
nucleon pairs and
quadrupole collective excitations that possess a clear symplectic
algebraic structure,
\spn{4} and \spR{3}, respectively.

The \Spn{4} dynamical symmetry of like-particle and proton-neutron
pairing correlations
\cite{Hecht,EngelLV96,SGD03} between nucleons occupying  the same
major shell has been
found to provide for a reasonable microscopic description of the
pairing-governed \IASs~
in light and medium mass nuclei \cite{SGD03stg,SGD04}. Currently,
these nuclei  have a
significant impact in astrophysical studies
\cite{Langanke98,Hosmer05} and have been extensively explored after
the advent of
radioactive beams. The comparison with realistic interactions can
determine the extend
to which the significantly simpler \Spn{4} model Hamiltonian  can
readily be used to
obtain an approximate, yet very good description of low-lying nuclear
structure and in
turn, one can apply the model to larger model spaces that are
otherwise prohibitive in
size. Furthermore, we introduce a possible
\Spn{4} symmetry breaking by an additional quadrupole-quadrupole
interaction and
examine the capacity of the extended model interaction to imitate realistic
interactions.  This, in turn, provides a further step towards gaining a better
understanding of the underlying foundation of the microscopic interactions.

\section{Symplectic \spn{4} Pairing Model Interaction}

The close interplay of like-particle and proton-neutron isovector
(isospin = 1) pairing
correlations have long been recognized as a major driver that  shapes
nuclear systems
with valence protons and neutrons occupying the same major shell.
While like-particle
pairing interactions are known to dominate far from the
$N=Z$ line, closer to it strong proton-neutron pair correlations are also very
important. Hence, isovector pairing plays a crucial role in understanding the
microscopic structure of light and medium mass nuclei around as well
as far off the
valley of  stability. A group-theoretical microscopic description of
isovector pairing,
based  on the fermion realization of the \so{5} algebra \cite{Hecht}
(isomorphic to
\spn{4}), was successfully applied to the structure of $fp$-shell $N=Z$ nuclei
\cite{EngelLV96}. These algebraic results have since been confirmed through
pairing-plus-quadrupole shell model work \cite{KanekoHZPRC59}.
Indeed, some recent
results  show that the symplectic \Spn{4} dynamical symmetry is
fundamental to the
nuclear  interaction that governs fully-paired \IASs~of light and
medium mass even-$A$
nuclei with valence protons and neutrons occupying the same shell
\cite{SGD04}.

The general model Hamiltonian with \Spn{4} dynamical symmetry for a
system of $n$
valence nucleons in a $4\Omega $-dimensional space consists of one-
and two-body
terms and can be expressed through the \Spn{4} group generators,
\begin{eqnarray}
\Hsp =&-G\sum _{i=-1}^{1}\hat{A}^{\dagger }_{i}
\hat{A}_{i}-F \hat{A}^{\dagger }_{0}\hat{A}_{0}-\frac{E}{2\Omega} (\hat{T}
^2-\frac{3\hat{N}}{4 })
\nonumber \\
&-D(\hat{T}
_{0}^2-\frac{\hat{N}}{4})-C\frac{\hat{N}(\hat{N}-1)}{2}-\epsilon
\hat{N},
\label{clH}
\end{eqnarray}
where $\hat{N}$ counts the total number of valence particles,
$\hat{T}^2=\Omega \{ \hat{T}_+,\hat{T}_-\}+\hat{T}_0^2$ is the
isospin operator,
$\hat{A}^{\dagger }_{0,+1,-1}$ creates a proton-neutron ($pn$) pair, a
proton-proton ($pp$) pair or a neutron-neutron $(nn)$ pair of total
angular momentum
$J^{\pi}=0^+$ and isospin $T=1$, $G,F,E,D$ and
$C$ are interaction strength parameters  and $\epsilon >0$ is the
Fermi level energy.
This Hamiltonian, which is rotationally invariant, conserves the
number of particles and
the third projection ($T_0$) of the isospin, while it includes
scattering of a $pp$ pair
and  a $nn$ pair into two  $pn$ pairs and vice versa, along with a
$J$-independent
isoscalar ($T=0$) $pn$ force.
The significant interplay between isovector and isoscalar interactions
is evident in the
low-lying structure of $N=Z$ odd-odd nuclei with valence protons and
neutrons filling the
same major shell.

Estimates for the interaction strength parameters in (\ref{clH}) were found
\cite{SGD03,SGD04} as a result from an optimal reproduction of the
Coulomb corrected
\cite{RetamosaCaurier} experimental energies
\cite{AudiWapstra,Firestone} of the lowest
isobaric analog
$0^+$ states of even-$A$ nuclei with valence nucleons occupying the
\flevel orbit or the
\fpg major shell \footnote{The lowest
\IASs~ of odd-odd nuclei have the isospin of the ground state of the
even-even same-mass
neighbor with a larger difference in proton and neutron  numbers.
These states are
ground states for even-even  nuclei and only some  [$N\approx Z$]
odd-odd nuclei.}. For
the $1f_{7/2}$ level with a $^{40}Ca$ core the  interaction strengths
were estimated to
be, $G/{\Omega }=0.453,\ F/{\Omega }=0.072,\ C=  0.473,\ D=0.149,\ E/{(2\Omega
)}=-1.120,\ \epsilon =9.359$. The analysis revealed that the model interaction
with \Spn{4} dynamical symmetry  accounts quite well for the
available experimental
energies of \IASs~ for a total of $149$ nuclei
\cite{SGD04} and in addition for the observed detailed structure
beyond mean-field
effects such as the $N=Z$ anomalies, isovector pairing gaps and
staggering effects
\cite{SGD03stg}. This in turn allowed us to interpret the main
driving force that
defines the properties of the states under consideration and to
provide a reasonable
description of these states, while retaining the physical validity
and the proper
limits of the strengths of interactions available in literature.

An important feature of our algebraic Hamiltonian (\ref{clH}) is that it arises
naturally within a microscopic picture. Because of this, the \Spn{4}
interaction can
be compared to realistic interactions and, as well, the physical
nature of the model
interaction and its strength can be realized. From a microscopic
perspective, the
pair-creation operators, $\hat{A}^{(\dagger )}$, and their annihilation counter
parts, $\hat{A}$, are realized in terms of creation $c _{jm\sigma}^\dagger$ and
annihilation $c _{jm\sigma }$ single-fermion operators with the standard
anticommutation relations
$\{c _{jm\sigma },c _{j^{\prime  }m^{\prime }\sigma ^{\prime }}^{\dagger
}\}=\delta _{j,j^{\prime }}\delta _{m,m^{\prime }}\delta _{\sigma ,\sigma
^{\prime }},$ where these operators create (annihilate) a particle of type
$\sigma =\pm 1/2$ (proton/neutron) in a state of total angular momentum $j$
(half integer) with projection
$m$ in a finite space $2\Omega =\Sigma _j (2j+1)$. There are ten independent
scalar products (zero total angular momentum) of the fermion operators:
\begin{eqnarray}
          \hat{A}^{\dagger }_{\mu }&=&
\frac{1}{\sqrt{2\Omega (1+\delta_{\sigma\sigma ^{\prime}})}}
\sum_{jm} (-1)^{j-m} c_{jm\sigma}^\dagger c_{j,-m,\sigma
^{\prime}}^\dagger,\nonumber \\
          \hat{A}_\mu &=& (\hat{A}^{\dagger }_\mu)^\dagger,\ \ (\mu
=\sigma+\sigma^{\prime})
\nonumber \\
\hat{T}_\pm &=& \frac{1}{\sqrt{2\Omega}} \sum_{jm} c^\dagger_{jm,\pm
1/2}  c_{jm,\mp 1/2}, \nonumber \\
\hat{N}&=& \sum_{\sigma jm} c^\dagger_{jm\sigma }  c_{jm\sigma },\
\hat{T}_0= \sum_{\sigma } \sigma \sum_{jm} c^\dagger_{jm\sigma }
c_{jm\sigma },
\label{gen}
\end{eqnarray}
which form a fermion realization of the symplectic \spn{4} Lie algebra. Such an
algebraic structure is exactly the one needed to describe isovector
(like-particle plus $pn$) pairing correlations and isospin symmetry
in nuclear \IASs.

Using relations (\ref{gen}), the one- and two-body interaction 
(\ref{clH}) can be rewritten
in standard second quantized form
in terms of fermion creation $a_{ jm(1/2)\sigma }^\dagger =c_{ jm(1/2)\sigma
}^\dagger$ and annihilation
$a_{ jm(1/2)\sigma } = (-1)^{j+m+1/2 +\sigma }c_{ j-m(1/2)-\sigma }$
tensor operators,
\begin{widetext}
\begin{eqnarray}
H&=&-\sum _{
{\scriptsize \begin{array}{c}
r\leq s \\
\{r=(j_r ,\half)\}
\end{array}}
}
\sqrt{[r]} \varepsilon _{rs} \{a_r^\dagger
\otimes a_s\}^{(00)}  -\sum_{
{\scriptsize \begin{array}{c}
r\leq s \\
t\leq u\\
\Gamma=(J,T)
\end{array}}
}
\frac{\sqrt{[\Gamma]}}{\sqrt{(1+\delta _{rs})(1+\delta _{tu})}}
W_{rstu}^\Gamma \{\{a_r^\dagger \otimes a_s^\dagger\}^\Gamma \otimes
\{a_t \otimes a_u\}^\Gamma \}^{(00)} \nonumber \\
&=&-\sum _{r\leq s}\sqrt{[r]} \varepsilon _{rs} \{a_r^\dagger
\otimes a_s\}^{(00)}  -\frac{1}{4}\sum_{
{\scriptsize \begin{array}{c}
rstu \\
\Gamma
\end{array}}
}
\sqrt{(1+\delta _{rs})(1+\delta _{tu})[\Gamma]} W_{rstu}^\Gamma
\{\{a_r^\dagger \otimes a_s^\dagger\}^\Gamma \otimes
\{a_t \otimes a_u\}^\Gamma \}^{(00)},
\label{V2ndQF}
\end{eqnarray}
\end{widetext}
with $[r]=2(2j_r+1)$ and $[\Gamma]=(2J+1)(2T+1)$, where
$\varepsilon _{rs}$ is the single-particle energy  and
$W_{rstu}^{JT} $ is the two-body antisymmetric matrix element in the
$JT$-coupled scheme [$W_{rstu}^{\Gamma
}=-(-)^{r+s-\Gamma}W_{srtu}^{\Gamma }=
-(-)^{t+u-\Gamma } W_{rsut}^{\Gamma }=(-)^{r+s-t-u}W_{srut}^{\Gamma }=
W_{turs}^{\Gamma }$].
For an isospin nonconserving two-body
interaction of isospin
rank ${\mathcal T}$, the coupling of fermion operators is as follows,
$\{\{a_r^\dagger \otimes
a_s^\dagger\}^{JT}\otimes \{a_t \otimes a_u\}^{JT } 
\}^{(0{\mathcal T})}$, with
$W_{rstu}^{({\mathcal T}) J T}$ matrix elements.
The latter are
expressed through the parameters of the model
interaction for
isospin rank $0$ and $2$ of $\Hsp$  and $\{r\le (s,t);\ t\le u\}$ 
orbits as follows,
\begin{eqnarray}
W_{rstu}^{(0)JT}&\equiv &W_{rstu}^{JT}=~<rsJTMT_0|H^{(0)}|tuJTMT_0>
\nonumber \\
&=&-(G+\frac{F}{3})\frac{\sqrt{\Omega _r \Omega _t}}{\Omega }\delta
_{(JT),(01)}
\delta_{rs}\delta_{tu} \nonumber \\
&-&\{-(\frac{E}{2\Omega}+\frac{D}{3})[(-)^T+\half]+C\}\delta_{rt}\delta_{su}
\label{W0me} \\
W_{rstu}^{(2)JT}&=&<rsJTMT_0|H^{(2)}|tuJTMT_0>
\nonumber \\
&=&\frac{\sqrt{2}}{3}(F\frac{\sqrt{\Omega _r \Omega _t}}{\Omega }\delta _{J0}\delta_{rs}\delta_{tu}
-D\delta_{rt}\delta_{su})\delta _{T1}.
   \label{W2me}
\end{eqnarray}
The isotensor part (\ref{W2me}) of the
model interaction introduces isospin dynamical  symmetry through the
$D$-term (retaining $T$ as a good  quantum number and splitting the
energy degeneracy
along the third projection of the  isospin) and as
well a plausible, but very weak, isospin mixing ($F$-term) \cite{SGD05}.

For the purposes of this paper, we will use only the isoscalar part
of our model
Hamiltonian (\ref{W0me}) and set all the orbits equal to $j=7/2$
($r=s=t=u$) because we choose to focus on a study of nuclei in the
single \flevel
level. In addition, these matrix elements correspond to the pure
nuclear interaction
and do not include Coulomb repulsion because its effect is corrected in the
experimental energies themselves by applying an empirical formula deduced in
\cite{RetamosaCaurier}. This may result in slightly more bound states predicted
by our model when compared to estimates of realistic interactions.

Within the isospin-invariant picture, the two-body matrix elements of the model
Hamiltonian $W_{rstu}^{JT}$ (\ref{W0me}) depend only on three
parameters, $ G_0=G+\frac{F}{3},\ E_0=
(\frac{E}{2\Omega}+\frac{D}{3})$ and $C$,
\begin{equation}
W_{\sevhalves \sevhalves \sevhalves \sevhalves}^{JT}
=-G_0 \delta _{(JT),(01)}-\{-E_0[(-)^T+\half]+C\}.
\label{W0s}
\end{equation}
The two-body matrix elements reflect the microscopic aspect of the model
interaction, which is $J$-independent for all but $J=0$. Hence \Hsp~
describes the average behavior of higher-$J$ states, while it
distinguishes between
$T=0$ and $T=1$ groups of states. The smaller the magnitude of $E_0$ ($<0$),
the smaller the separation of these groups. As expected, the pairing
correlations
contribute only to the $(J=0,T=1)$ state and they are absent for
higher-$J$ states
where both particles are uncoupled. Relative to the $0^+$ $T=1$
state, the bigger the
$G_0$ pairing strength, the larger the energy gap to the
higher-$J$ states.

The role of the \Spn{4} dynamical symmetry in generating the
energy spectrum of the \flevel nuclei can be further understood by comparing
the \Spn{4} interaction to the \CDB \cite{MachleidtSS96M01}, 
\CDBt~\cite{PopescuSVN05} and
\Gm~\cite{HonmaOBM04} realistic interactions. \CDB~is a charge-dependent
one-boson-exchange nucleon-nucleon ($NN$) potential that is one of 
the most accurate
in reproducing the world proton-proton and neutron-proton scattering data.
In addition, the \CDBt~interaction introduces phenomenological
isospin-dependent central terms plus a tensor force with strengths 
and ranges determined in
no-core $0\hbar\omega $ shell model calculations to achieve an 
improved description of the
$A=48$ Ca, Sc and Ti isobars. The \Gm~ effective interaction is 
obtained from a realistic
G-matrix interaction based  on the Bonn-C potential \cite{Gint} by 
adding empirical
corrections determined through systematic fitting to experimental 
energy data in the
$fp$ shell.

\section{Theory of Spectral Distributions}
Group theory underpins spectral distribution theory
\cite{FrenchR71,ChangFT71,HechtDraayer74,Parikh78,Kota79}. The model space is
partitioned according to particular group symmetries and each
subsequent subgroup
partitioning yields finer and more detailed spectral estimates. For
$n$ particles
distributed over ${\mathcal N}$ single-particle states, a {\it
scalar} distribution
(denoted by ``$n$" in the formulae) is called the spectral
distribution averaged
over all $n$-particle states associated with the \Un{}{\dimN=4\Omega} group
structure and an {\it isospin-scalar} distribution (denoted by
``$n,T$") is averaged
over the ensemble of all
$n$-particle states of isospin $T$ associated with $\Un{}{\dimN=2\Omega}
\otimes \Un{}{2}_T$.

For a spectral distribution $\alpha $ ($\alpha $ is $n$ or $n,T$), the
correlation coefficient between two Hamiltonian operators,
$H$ and $H^\prime$, is defined as
\begin{eqnarray}
\zeta ^\alpha _{H,H^\prime }&=&\frac{\langle (H^\dagger -\langle
H^\dagger \rangle
^\alpha ) (H^\prime-\langle H^\prime \rangle ^\alpha )
\rangle ^\alpha }{\sigma _H \sigma _{H^\prime}} \\
&=&\frac{\langle H^\dagger H^\prime\rangle ^\alpha -\langle H^\dagger \rangle
^\alpha \langle H^\prime \rangle ^\alpha }{\sigma _H \sigma _{H^\prime}},
\label{zeta}
\end{eqnarray}
where the ``width" of the distribution is the positive square root of the
variance,
\begin{eqnarray}
(\sigma ^\alpha _{H})^2=\langle (H-\langle H\rangle ^\alpha )^2
\rangle ^\alpha
=\langle H^2 \rangle ^\alpha -(\langle H \rangle ^\alpha )^2.
\label{sigma}
\end{eqnarray}
The average values, related
to the trace of an operator divided by the dimensionality of the
space, are given in
terms of the ensemble considered.
In the (isospin-)scalar case, the correlation will be denoted by $\zeta ^n$
($\zeta ^{n,T}$) or simply $\zeta $ ($\zeta ^T$) for $n=2$.

The steps for computing the $\zeta ^{\alpha }$ correlation 
coefficient and the $\sigma
^{\alpha }$ variance
\footnote{This follows from the decomposition of the 
one($k=1$)- and two($k=2$)-body
interaction $H$ into definite particle rank terms
[irreducible tensors ${\mathcal H}_k(\nu )$ of rank $\nu =0,1,2$], 
that is into a collection
of pure zero-, one- and two-body interactions.  For example, in the 
scalar single-$j$ case
for $n$ particles, the Hamiltonian can be rendered,
$
H=n{\mathcal H}_1(0)+\binom{n}{2} {\mathcal H}_2(0)+{\mathcal
H}_1(1)+(n-1){\mathcal H}_2(1) +{\mathcal H}_2(2)
=-n\varepsilon -
\binom{n}{2}W_c  -\frac{1}{2}\sum_{\Gamma}
\sqrt{[\Gamma]}
W_{rrrr}^\Gamma (2)\{\{a_r^\dagger \otimes a_r^\dagger\}^\Gamma \otimes
\{a_r \otimes a_r\}^\Gamma \}^{(00)},
$
where $\varepsilon $ is the single-particle energy and for a single-$j$
level the pure one-body part is trivially zero.}
are given in
\cite{FrenchR71,ChangFT71,French72,HechtDraayer74,Kota79}  (see also 
computational codes
\cite{Kota79,ChangDW82}) and take on the simple form for a single-$j$ level:
\begin{widetext}
\begin{eqnarray}
\langle H^\dagger H^\prime\rangle ^\alpha -\langle H^\dagger \rangle
^\alpha\langle
H^\prime \rangle ^\alpha
&&=\sum _\tau  p_2(\alpha ,\tau )\frac{1}{\sum_{\Delta}[\Delta]}
\sum_{J}
[\Delta] W_{rrrr}^{J\tau } (2) {W^\prime }_{rrrr}^{J\tau } (2),
\label{<HH'>} \\
W_{rrrr}^{J\tau }(2)&&=W_{rrrr}^{J\tau }-W_c^{(\tau )}, \\
W_c^{(\tau )}&&= \frac{1}{\sum _{\Delta}[\Delta]}\sum 
_{\Delta}[\Delta] W_{rrrr}^{J\tau },
\label{Wc}
\end{eqnarray}
where $\tau =\{0 \ \text{or } 1 \}$ is the isospin label of the 
two-body matrix elements,
$W_{rrrr}^{J\tau }(2)$ is the traceless {\it pure two-body} 
interaction and $W_c^{(\tau )}$
is the monopole moment or centroid in the (isospin-)scalar case, that 
is the average
expectation value of (the isospin-$\tau $ part of) the two-body 
interaction for a
two-particle system $n=2$.
In the scalar ($\alpha =n$) case the following holds,
$\Delta =\Gamma =(J,\tau )$,
$\sum_{\Gamma}[\Gamma]=\binom{\mathcal N}{2}$, ${\mathcal N}=4\Omega
=2(2j_r+1)$, and the $\tau $-
independent propagator is
\begin{eqnarray}
p_2(n ,\tau )=\frac{n(n-1)({\mathcal N}-n)({\mathcal N}-n-1)}
{2({\mathcal N}-2)({\mathcal N}-3)}.
\end{eqnarray}
In the isospin-scalar ($\alpha =n,T$) case:
$\Delta =J $,
$\sum_{J}[J]=\frac{{\mathcal N} ({\mathcal N}+(-1)^\tau )}{2}$, 
${\mathcal N}=2\Omega $, and  the propagator
functions are \cite{French69,HechtDraayer74}
\begin{eqnarray}
p_2(n,T,\tau =0)&=& \frac{[n(n+2)-4T(T+1)]
[(\dimN -\frac{n}{2})(\dimN -\frac{n}{2}+1)-T(T+1)]}{8\dimN (\dimN-1)} \\
p_2(n,T,\tau =1)&=& \frac{1}{\dimN (\dimN +1)(\dimN -2)(\dimN -3)}
\{\half T^2 (T+1)^2(3\dimN ^2-7\dimN +6) \nonumber\\
&&+\frac{3}{8}n(n-2)(\dimN -\frac{n}{2})(\dimN -\frac{n}{2}+1)(\dimN
+1)(\dimN +2)
\nonumber\\
&&+\half T(T+1)[(5\dimN -3)(\dimN +2)n(\frac{n}{2}-\dimN) +
\dimN (\dimN -1)(\dimN +1)(\dimN +6)]\}.
\end{eqnarray}
\end{widetext}

In terms of a geometrical picture, the correlation coefficient $\zeta $
defines the angle between two vectors ($H$ and $H ^\prime$) of
length $\sigma _{H^{(\prime )}}$ (\ref{sigma}) and hence its square gives a
normalized measure (percentage) of one of the vectors, e.g. the \Spn{4}
interaction, that is contained in the other, such as a realistic
interaction. The correlation coefficient is a measure that is
independent of the averages
of the interactions. Clearly these averages, though an interesting measure, are
irrelevant when the focus is on detailed property-defining two-body
interaction beyond
strong mean-field effects.

For the \Spn{4} interaction, the average two-body interaction is
expressed in terms of the model parameters in the scalar case as,
\begin{equation}
W_c=-\frac{3G_0}{\binom{\dimN}{2}}+\frac{3E_0}{2(\dimN -1)}-C
\end{equation}
and in the isospin-scalar case as
\begin{equation}
W_c^T=-\frac{G_0}{\binom{\dimN}{2}}\delta _{T1} +E_0[(-1)^T+\half ]-C.
\end{equation}
Hence the pure two-body $W_{\sevhalves \sevhalves \sevhalves
\sevhalves}^{JT}(2)$ matrix elements  (\ref{W0s}), and consequently 
the correlation
coefficients involving \Hsp, are independent of the $C$ (and $E_0$) 
parameter(s) in the
(isospin-)scalar case.

\section{Underlying Symmetries of Realistic Interactions}

We now use statistical concepts to probe the nature of the
\CDB \cite{MachleidtSS96M01}, \CDBt~ \cite{PopescuSVN05} and
\Gm~\cite{HonmaOBM04} realistic interactions, hereafter referred as
$H_R$. Specifically,
we  will compare these interactions to the symplectic pairing and quadrupole
interactions through their mutual correlations. Clearly, if two
interactions have
similar matrix elements they will be strongly correlated and any
pattern that is
observed in the behavior of one will be reflected in the other. This
can be made
quantitative by evoking measures from statistical spectroscopy,
namely, the closer  the
correlation coefficient between two interactions is to unity the more
similar  their
spectra with the two coinciding within a rescaling factor when the correlation
coefficient is unity.

In a similar manner, the projection onto a model Hamiltonian that
describes collective
rotational excitations or/and pairing correlations can be used to
probe the rotational
and pairing  characteristics of a microscopic interaction
\cite{Draayer73,HalemaneKD78,KotaPP80,CounteeDHK81}. The dynamical
symmetry of the
pairing (or quadrupole-quadrupole) interaction sets a specific
relation between the
matrix elements of the Hamiltonian that models it. If this relation
is found in a
realistic interaction, that is, the model and realistic interactions
are strongly
correlated, then the latter possesses the underlying symmetry and
will reflect the
characteristic properties of the pairing (quadrupole) Hamiltonian. It
should be clear that
the complement is also true, namely, if a model interaction is
strongly correlated with a
realistic one, the associated model calculations can be used to
investigate the behavior of
physical systems.

\subsection{The \Spn{4} Model and Pairing Character}
An interesting feature of any interaction is its trace-equivalent 
part. If the latter is
found dominant then only the underlying group scalars are enough to 
provide for an
approximate and yet reasonably good solution
\cite{FrenchR71}. The greatest advantage in this case is the 
simplicity of the many-body
problem and the tractable size of the model space. In the 
isospin-scalar case, the centroid
of a Hamiltonian expressed through the $\varepsilon$ 
single-particle energy and the $W_c^{0,1}$ monopole moments
(\ref{Wc}) is \cite{French69}
\begin{widetext}
\begin{eqnarray}
\langle H^\dagger \rangle ^{n,T} =-n\varepsilon -
\binom{n}{2}\frac{W_c^0+3W_c^1}{4} -[T(T+1)-\frac{3}{4}n]\frac{W_c^1-W_c^0}{2}.
\end{eqnarray}
\end{widetext}
For a Hamiltonian with symplectic dynamical symmetry, \Hsp, the
trace-equivalent part in the isospin-scalar distribution includes the
$E$-, $C$- and
$\epsilon$-terms of (\ref{clH}). When applied to the lowest
\IASs~ of the nuclei in the \flevel orbit, it reproduces their energy
within 1\% of
the experimental value for about a third of the nuclei. While for
these states the
centroid is sufficient to achieve a good description, its difference with
experiment goes up to 7\% compared to only 0.4\%  achieved by the whole \Spn{4}
Hamiltonian. In addition, a model with a  trace-equivalent \Spn{4}
Hamiltonian will
not be capable of explaining the fine nuclear structure where \Hsp~
(\ref{clH}) succeeded \cite{SGD03stg} and will not correlate with any of the
realistic interactions. The latter indicates an inadequate
reproduction of the entire
energy spectrum. Indeed, while such an interaction was found insufficient for a
description of $ds$ shell nuclei when compared to several effective
interactions, a
drastic improvement was achieved with the inclusion of pure two-body residual
interactions of the spin-orbit and quadrupole types
\cite{HalemaneKD78} as well as pairing correlations \cite{CounteeDHK81}.
In summary, the \Spn{4} symmetric Hamiltonian (\ref{clH}) provides for a more
accurate description of nuclear structure by adding to an average interaction
suitable for the \IASs~ in \flevel a significant isovector pairing part.

Furthermore, the \Spn{4} dynamical symmetry allows the model 
Hamiltonian to reflect on the charge dependence of
the nuclear interaction, which is evident from 
experiments and present in almost
all of the modern realistic interactions (e.g., \CDB). While the 
small isospin admixture found in the \flevel \IASs~ has
been directly estimated through the \Hsp~ eigenstates \cite{SGD05}, 
the theory of
spectral distributions provides a further estimate of isospin 
symmetry breaking throughout the entire spectrum
\cite{French69} based on the \Spn{4} isotensor interaction 
(\ref{W2me}). Using equations (4), (15) and (25) in
\cite{HechtDraayer74} the isospin $T+2$ admixture into an average $T$ 
state is found in \flevel to be, as expected, much
smaller (on average less than $0.0001\%$) than the one detected among 
the $0^+$ seniority-zero states \cite{SGD05}. As
expected, it is also much smaller than the measure calculated in 
\cite{HechtDraayer74} for the two-body Coulomb
interaction in the \flevel shell (with a maximum value of $0.009\%$), 
because the latter corresponds to the stronger
$\Delta T=1$ admixture. The quite small isospin symmetry breaking 
that the isotensor \Spn{4} model interaction
introduces allows us to carry the present study without its consideration.

The extent to which the \Spn{4} dynamical symmetry governs the $H_R$ realistic
interactions within a certain domain of states is represented by the
correlation
coefficients between \Hsp~ and $H_R$ (Table \ref{tab:CC0}).
\begin{table}[th]
\caption{Correlation coefficients for a two-nucleon system, $n=2$, in
the scalar
($\zeta $) and isospin-scalar ($\zeta ^{T}$) distributions.}
\begin{tabular}{cccc}
\hline
$\zeta $         & \CDB        & \CDBt       & \Gm \\
$(\zeta ^{T=0},\zeta ^{T=1}) $ & &             &   \\
\hline \hline
\Hsp           &   0.66         &  0.64              &  0.76  \\
                  &(\, - \, ,0.61) & (\, - \, ,0.85)    & (\, - \,,0.71) \\
\CDB           &                &  0.95              &  0.96  \\
                  &                & (0.99,0.94)        & (0.98,0.99) \\
\CDBt          &                &                    &  0.97  \\
                  &                &                    & (0.99,0.97) \\
$H_Q^\perp(2)$ &   0.47         &  0.60              &  0.53  \\
                  &  (0.60  ,0.73) &   (0.68  ,0.50)    & (0.74,0.65) \\
$H_M$          &   0.81         &  0.87              &  0.93  \\
                  &  (0.60  ,0.95) &   (0.68  ,0.98)    & (0.74,0.96) \\
\hline
\end{tabular}
\label{tab:CC0}
\end{table}

In the {\bf scalar} distribution, where the outcome is averaged over
the isospin
values, the analysis of the results shows that all of the   realistic
interactions
correlate between themselves to a high degree (Table
\ref{tab:CC0}, upper cell for each pair of interactions). In
comparison, each of them
has a correlation with the \Spn{4} symmetric interaction of  order of
0.6-0.8, which is
typically regarded as a good one \cite{Rosensteel80}. This  implies
that the realistic
interactions possess around 40-60\% of the dynamical symplectic
symmetry of \Hsp~
[$(\zeta _{\Hsp,H_R})^2$] and hence 0.4-0.6 portion of $H_R$ is
dynamically symmetric
under \Spn{4} transformations. Equivalently, the \Hsp~ model
interaction contains 40-60\%
of the realistic interactions under consideration. This is a very
interesting result,
and definitely valuable concerning the restrictions the symplectic
model is subject to.

A much more interesting scenario occurs when the {\bf isospin-scalar} case is
considered. This is because in this case the space is divided into  two regions
specified by their isospin values with a view towards a more detailed
examination of
the nature of the interactions under consideration. Indeed the
centroids of both $T=0$
and $T=1$ regions are considerably separate as is observed in the realistic and
\Spn{4} interactions and as well confirmed by experiment. In
addition, the important
pairing correlations that are described in the symplectic model enter
in the $T=1$
channel, where the tendency towards  pair formation of realistic
interactions can be
detected.

As in the scalar case, all of the realistic interactions are quite
strongly correlated
in both the $T=0$ and $T=1$ channels (Table \ref{tab:CC0}, lower cell
for each pair of
models). The $T=1$ correlation coefficients between \Hsp~ and the
$H_R$ realistic
interactions do not depend on any of the parameters in \Hsp~
including the pairing
strength itself
\footnote{For the isospin-scalar distribution, the \Spn{4} $T=0$
model interaction
contains only a trace-equivalent part and hence the correlation
coefficients cannot be
determined.} and show that  \Hsp~ correlates strongly with the
realistic interactions.
Among the three $H_R$, the $T=1$ part of the \CDBt~ interaction
possesses the closest
similarity to the $pn$ and like-particle $J=0$ pairing correlations.
This is indicated
by its large projection of 72\% [$(\zeta ^1_{H_{\spn{4}},H_R})^2$ in Table
\ref{tab:CC0}] onto the $T=1$
\Hsp~ pairing interaction. Hence, the \CDBt~ interaction is expected
to describe quite
well phenomena of a pairing character.

The individual pairing strength associated with each realistic
interaction is typically
invoked for purposes of comparison. Compared to \CDB, the $J=0$
isovector pairing
strength estimate turns out to be stronger for both
\CDBt~ and
\Gm~ with a relatively weaker coupling observed in the latter.
Indeed, \Gm~ was shown to
tend towards smaller $J=0$ pairing strength
\cite{HonmaOBM04} when compared to two other effective interactions,
namely the $G$
interaction based  on the Bonn-C potential \cite{Gint} and the KB3G interaction
\cite{KB3G}.
However, pairing effects, with strong or weak coupling, may be fully
or partially
reflected in $H_R$ considerably depending on the strength of the
overall interaction. It
is the correlation coefficient between a realistic ($H_R$) and
symmetry-holding (as
\Hsp) interactions that manifests what part of $H_R$ is ruled by the
symmetry.  In this
sense, we can identify that pairing  features are more fully
developed in \CDBt, then in
\Gm~ and the least in \CDB~ ($\zeta ^1_{H_{\spn{4}},H_R}$ in Table
\ref{tab:CC0}). In
the latter, other types of interaction compete stronger with pair
formation than in the
former  two interactions and hence suppress pairing coherence.

The large $J=0$ coherence and its strong coupling observed in the
\CDBt~ interaction
should not be surprising because it reproduces (by an optimal fit) the energy
difference between the ground state and the first $2^+$ state of
$^{48}$Ca. Such an observable is believed to be directly affected by
the formation of
correlated pairs in the ground state of the spherical core of
$^{48}$Ca and the pairing
gap that occurs below the first excited state of a broken pair. It is
interesting to
point out that while the addition of three phenomenological terms  to the \CDB~
interaction to obtain \CDBt~ keeps the close similarity between both
interactions
[$\zeta ^{1}=0.94$ (Table \ref{tab:CC0})], it causes the
correlations with isovector pairing interaction to double in strength [$(\zeta
^1_{H_{\spn{4}},H_R})^2$].  In short, the
analysis shows that the simple \Spn{4} model interaction can
reproduce reasonably the
$T=1$ low-lying energy  spectra generated by the \CDBt~ realistic
interaction for a
system of two nucleons in the
\flevel orbit  and for this reason can be used as a good approximation.

Another result in favor of the algebraic \spn{4} model follows from a
comparison of the
lowest \IASs~ in the $A=42$ isobars. These are precisely the
$(J=0,T=1)$ states, which are expected to be shaped by strong
proton-neutron and
like-particle pairing correlations \cite{BohrMottelsonPines} and are
well described by
the \spn{4} model \cite{SGD04}. The outcome reveals a very close
similarity between the
estimate of the two-body $(J=0,T=1)$ matrix element for the
symplectic interaction
($-1.85$ MeV) and both \CDBt~ ($-2.06$ MeV) and \Gm~ ($-2.44$ MeV) realistic
interactions.  In addition, the energy differences between the first
$2^+$ state and the
$0^+$ ground  state for the different effective interactions are also
very close,
namely, $1.91$ MeV (for \Hsp),
$2.00$ MeV (for \CDBt) and $1.50$ MeV (for \Gm). All these estimates are rather
different from \CDB~ with $0.48$ MeV $2^+$ to $0^+$ energy difference
but very close to
the experimental energy gap for the $A=42$ isobars, namely,
$1.56$ MeV for $^{42}$Ti, $1.59$ MeV for $^{42}$Sc and $1.52$ MeV for
$^{42}$Ca.

\subsection{Pairing $+$ Quadrupole Model Interaction}

While the pairing-governed \IAS~ energies are well determined within 
the framework
of
the \Spn{4} model, the nuclear spectrum as described by the
\Spn{4}-symmetric  Hamiltonian contains degenerate higher-$J$ states
averaged for a
given isospin value as can be clearly seen from its microscopic
structure (\ref{W0me}).
Nonetheless, the correlation of \Hsp~ with realistic interactions for
the \flevel level
turns out to be reasonably strong.  A question one can pose concerns
the role of other
significant interactions in  nuclei such as the quadrupole-quadrupole
interaction ($Q
\cdot Q$). As our results  indicate, answers to such questions can be
found within the
framework of statistical measures.

The pairing model based on the \spn{4} algebra [incorporating
like-particles pairing
through an \su{}{2} subalgebra] is commonly considered to be
inappropriate for two
reasons. The first reason is related to the degeneracy of the
single-particle levels,
which is not a problem for the \flevel shell considered as a  single orbit
well-separated from the $ds$ shell and the upper $fp$ shell. The
second reason is the
lack of the $Q \cdot Q$ interaction. This is because one usually
neglects the fact that
the pairing interaction contains in itself a part of the quadrupole-quadrupole
interaction. This part is not negligible with  the correlation being
typically between
0.4 to 0.6 depending on the distribution  considered: 15\% when the
whole space is
considered and 35\% in the $T=1$ region. This is  probably one of the
reason why the
\Spn{4} model interaction turns out to work rather  well despite an
explicit appearance
of the quadrupole-quadrupole interaction.

Because of the fact that the $Q \cdot Q$ interaction is already present in the
\spn{4} Hamiltonian, its additional influence can be studied following the
construction prescribed in \cite{HalemaneKD78}. In short, we add a $Q
\cdot Q$ term
to the symplectic \Hsp~ Hamiltonian (\ref{clH}) in a way that this term is
only the part of the pure two-body quadrupole-quadrupole interaction
that is not
contained in \Spn{4}, or in the vector algebra terminology we add only the
part that is orthogonal to the pure two-body \Spn{4} Hamiltonian 
\cite{Potbhare77},
\begin{equation}
H_M=H_{\spn{4}}+H_Q^\perp(2),\
H_Q=-\frac{\chi }{2} Q \cdot Q .
\label{HM}
\end{equation}
Such a Hamiltonian does not affect the centroid of \Hsp~ because
$H_Q^\perp (2)$ is traceless. In this way this collective interaction
preserves the
shell structure that is built into \Hsp~ and established by a
harmonic oscillator
potential and as a result is favored in many studies
\cite{HalemaneKD78,CounteeDHK81,DraayerR83b,BahriDCR90}.

Compared to the pairing \Hsp~ Hamiltonian, the additional collective
interaction,
$H_Q^\perp (2)$, has a lower  correlation with $H_R$ for all the
cases except for $T=1$
\CDB~ and where $\zeta _{H_{\spn{4}},H_R}$ cannot be determined (Table
\ref{tab:CC0}). The realistic interactions contain the \Spn{4}
interaction by 5\% to 50\% more than they contain $H_Q^\perp (2)$, 
the largest value
being for the $T=1$ \CDBt~ interaction.
We should emphasize that this outcome does not imply that the
realistic interactions
correlate better with the pairing interaction  than they do with the
quadrupole-quadrupole interaction nor that their pairing character is
dominant. This is
because the $H_Q^\perp (2)$ interaction represents  only that part of
the rotational
interaction that is not included in the
\Spn{4} interaction and the entire $Q \cdot Q$ collective mode
affects both $\zeta
^{(T)}_{\HQ,H_R}$ and $\zeta^{(T)}_{H_{\spn{4}},H_R}$ correlations.
The outcome only
implies that a comparatively larger part of the overall  correlations
is already
accounted for solely by the symplectic \spn{4} algebraic  model interaction.

In our study, we vary only $\chi$, the quadrupole strength parameter 
in (\ref{HM}), to find
its optimal value (which is an exact solution) by maximizing the 
correlation coefficient
$\zeta$ between $H_M$ and $H_R$ \cite{Chang78}. We do not alter the 
parameters of the
\Spn{4} model, which have already been shown in an appropriate domain 
of states to be valid
for reproducing various quantities (such as binding energies and 
pairing gaps) and are in
agreement with estimates available in literature 
\cite{SGD03stg,SGD04}. This implies that
the $\sigma $ ``width" of \Hsp~ (\ref{sigma}) does not change. The 
minimization procedure is
performed for $H_M$ compared to each realistic interaction and in the 
isospin-scalar case,
for each isospin value (Tables
\ref{tab:CC0} and \ref{tab:enMoments}).
\begin{table}[th]
\caption{First and second energy moments [the centroid $W_c$
(\ref{Wc})
and the ``width" $\sigma $ (\ref{sigma})] of realistic and model interactions
for a two-nucleon system, $n=2$, in the \flevel level. $H_M$ is
determined by an estimate for
the quadrupole-quadrupole strength $\chi$ for each realistic interaction; its
centroid energies coincide with the ones of \Hsp~ for a given distribution.}
\begin{tabular}{cccc}
\hline
         & \CDB        & \CDBt       & \Gm \\
\hline \hline
\multicolumn{4}{c}{Scalar Distribution} \\
$W_c$            & 0.30   & 0.19   & -0.60   \\
$\sigma $        & 0.59   & 0.80   & 0.98    \\
$H_M,\ W_c$      & -0.63  & -0.63  & -0.63   \\
$H_M,\ \sigma$   & 1.23   & 1.36   & 1.21    \\
$H_M,\ \chi$     & 0.096  & 0.124  & 0.092    \\
\multicolumn{4}{c}{Isospin-scalar Distribution, $T=1$} \\
$W_c^1$          & 0.54   & 0.46   & -0.17    \\
$\sigma ^1$      & 0.27   & 0.57   & 0.62    \\
$H_M,\ W_c^1$    & -0.01  & -0.01  & -0.01   \\
$H_M,\ \sigma^1$ & 0.55   & 0.41   & 0.48    \\
$H_M,\ \chi$     & 0.071  & 0.036  & 0.055   \\
\hline
\end{tabular}
\label{tab:enMoments}
\end{table}

\begin{figure}[th]
\centerline{\epsfxsize=1.6in\epsfbox{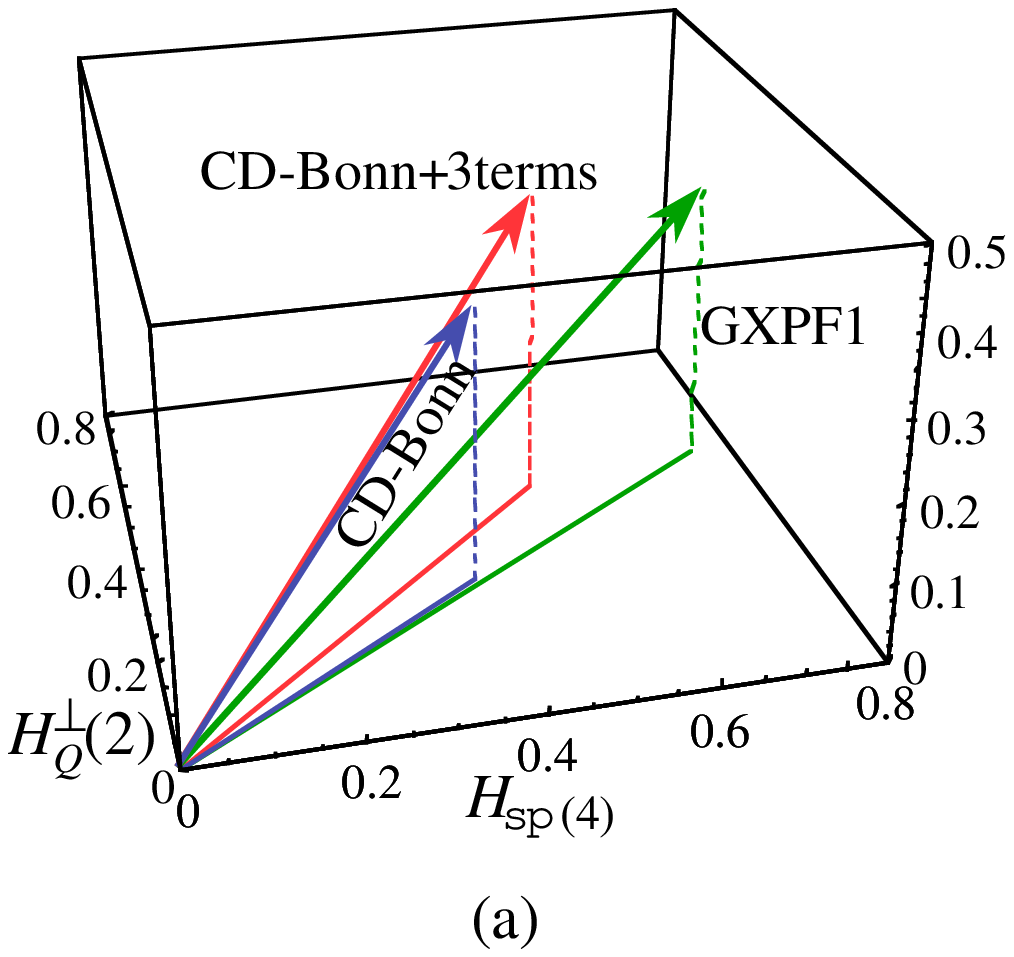}
\epsfxsize=1.6in\epsfbox{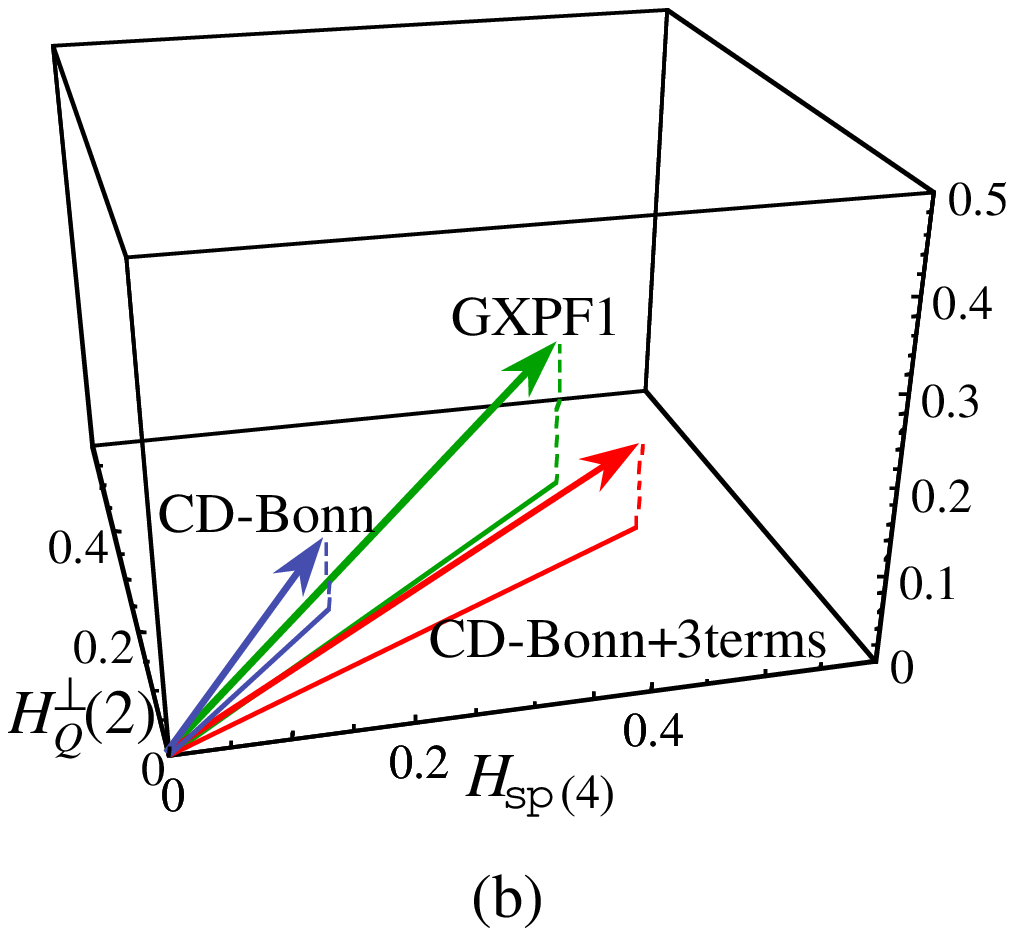}}
\caption{(Color online) Geometrical representation of the realistic 
interactions,
\CDB~ (light blue),
\CDBt~(red) and \Gm~(green), in an abstract operator space, where the
horizontal
plane is spanned by the orthogonal linear operators, the pure
two-body \Hsp~ and
$\HQ$ model Hamiltonians, both linearly independent of additional operators
represented by the vertical axis. (a) Scalar distribution. (b)
Isospin-scalar distribution, $T=1$. The orientation of the
vectors remains the same
for any particle number $n \ge 2$ in (a) and for all $T=n/2$ cases in (b).}
\label{Vs}
\end{figure}

In both scalar and isospin-scalar cases, the addition of the
quadrupole-quadrupole interaction definitely improves the
$\zeta^{(T)}_{H_{M},H_R}$
correlation (Table
\ref{tab:CC0}), which is associated with the angle between the
two-body effective
interaction and its projection on the plane spanned by two orthogonal
vectors, two-body
\Hsp~ and $H_Q^\perp(2)$, in an abstract operator space (Figure \ref{Vs}).
In the same representation, the angles between the realistic
interactions and both
axes give the
$\zeta ^{(T)}_{\HQ,H_R}$ and $\zeta^{(T)}_{H_{\spn{4}},H_R}$ correlations, and
the length of each vector is specified by $\sigma ^{(T)}_H$ (\ref{sigma}).
Therefore, while enhanced quadrupole effects rotate the
projection of an interaction closer to the $\HQ $ axis, greater influence of
additional interactions neglected in $H_M$ pushes $H_R$ away from the $H_M$
horizontal plane.

The \Gm~ interaction is found to correlate best with the model $H_M$
Hamiltonian for the scalar distribution (86\%) and in the $T=0$ case
compared to the
other interactions (Table \ref{tab:CC0}). More than 50\% of the $T=0$ \Gm~
interaction is accounted by the isoscalar model interaction. The
$T=0$ correlation
coefficients between the model $H_M$ Hamiltonian and $H_R$ are
reasonably good and
determined solely by the $Q \cdot Q$ interaction independent of its
strength, $\chi
$ (\ref{HM}).

In the $T=1$ region,  all the realistic interactions considered are
reproduced to
the 90\% - 97\% level by such a pairing$+$quadrupole model interaction.
Other interactions contribute almost negligibly as is clearly seen from
Figure \ref{Vs} (b) with their contribution being least for the
\CDBt~ interaction.
In addition, within the $T=1$ distribution, the smallest $\chi $ value
(Table \ref{tab:enMoments}) is found for the \CDBt~ realistic interaction as
expected due to its reasonable correlation with the
\Spn{4} model interaction. In summary, the results once again
prove that the pairing and quadrupole-quadrupole interactions are
significant in
shaping nuclear structure and are dominant for the
$T=1$ two-body nuclear interaction for the \flevel orbit.

  From the point of view of the model interaction (\ref{HM}) we
adopted, the rotational
character of the three realistic interactions may appear to be
obscure because a strong
correlation to the entire $H_Q$ quadrupole-quadrupole  interaction
(not only to its
projection \HQ),
$\zeta ^{(T)}_{H_Q,H_R}$, is needed. However, development of rotational
features turns out to follow qualitatively the $\zeta
^{(T)}_{\HQ,H_R}$ correlation
coefficients (Table \ref{tab:CC0}). This is
because $H_Q$ is already present in the $H_M$  model Hamiltonian and
one can visualize $H_Q$ in the scalar (isospin-scalar, $T=1$) case in Figure
\ref{Vs} as an axis that lies in  the horizontal plane around 67${^\circ}$
(54${^\circ}$) counterclockwise from the \Hsp~ axis and hence it lies
furthest away
from \Hsp~ compared to all three realistic interactions. Therefore,
the comparatively
largest collectivity is attained within
\CDBt~ in the  scalar case, within \Gm~ for the $T=0$ domain of
states (with an exact
quantitative measure
$\zeta ^{0}_{H_Q,H_R}=\zeta ^{0}_{\HQ,H_R}$), and within \CDB~ in the
$T=1$ case. In contrast, the $T=1$ part of the \CDB~ interaction
shows the  smallest
individual quadrupole strength, which is an illustrative example of a
very  prominent
rotational behavior (detected via $\zeta $) but of a weak strength
(depending on $\sigma $). In general, the individual quadrupole 
strength associated with the
\Gm~ interaction is very similar to, yet slightly stronger than, the
one of \CDBt~ with
the same trend observed for \Gm~\cite{HonmaOBM04} with respect to KBG3.

\subsection{Energy spectrum}
The scalar and isospin-scalar $T=1$ distributions show strong
correlations of the
pairing$+$quadrupole model interaction with the realistic
interactions  and hence a
similar pattern of energy states is expected.
\begin{widetext}
\center{
\begin{figure}[th]
\centerline{\epsfxsize=5.7in\epsfbox{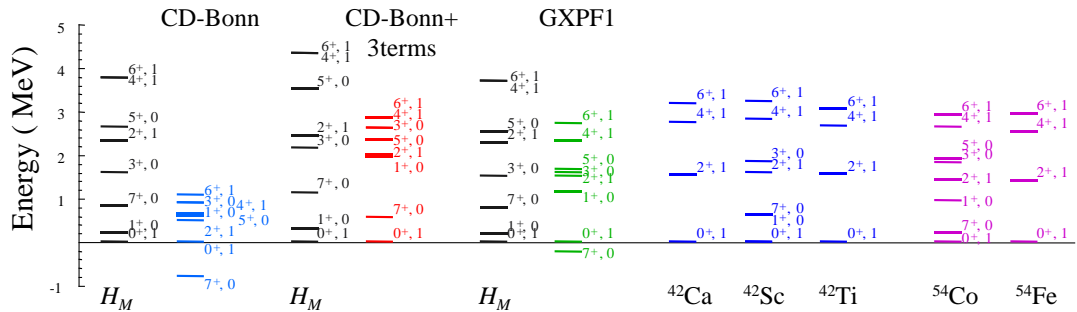}}
\caption{(Color online) Energy spectra of two-particle states in the 
\flevel level
predicted by the
\CDB~ (light blue), \CDBt~ (red) and \Gm~ (green) realistic
interactions. Each is
compared to the model Hamiltonian $H_M$ (black) with $\chi =0.096,\ 0.124$ and
$0.092$, respectively (Table \ref{tab:enMoments}). For comparison, the
available experimental
energy spectra of the $A=42$ Ca, Sc, Ti isobars (blue) and $A=54$ Co
and Fe isobars (magenta)
are  also shown.}
\label{scalarSpectra}
\end{figure}
}
\end{widetext}

While correlation coefficients
(\ref{zeta}) prove useful in studies of nuclear properties shaped by
the residual pure
two-body interaction, the discrete energies of a  quantum-mechanical system are
additionally influenced by the centroid, $W_c^{(T)}$
(\ref{Wc}), and the
overall interaction strength related to $\sigma ^{(T)}$
(\ref{sigma}). The centroid of
the $H_M$ pairing$+$quadrupole model interaction, which coincides
with the one for
\Spn{4} ($W^{(T)} _{c,H_M}=W^{(T)} _{c,H_{\spn{4}}}$) (Table
\ref{tab:enMoments}), is very close to that for \Gm~ and both differ
from the other two
realistic interactions. However, this quantity is irrelevant for the
energy spectra
relative  to the ground state within a given distribution.

For the scalar distribution, the $H_M$ model Hamiltonian generates
energy spectra that
are comparatively more spread out, especially with respect to
\CDB~ (see Figure
\ref{scalarSpectra} and $\sigma $ in Table \ref{tab:enMoments}). A
renormalization of
$H_M$ will push the higher-lying states down and will establish an
energy pattern very
much like the ones observed in \Gm, \CDBt~ and  experiment (except for the
$1^+$ state). Such a renormalization, however, is not done because the scalar
distribution itself introduces averaging over isospin values. Instead
we turn to the
more detailed $T=1$ spectral distribution.

\begin{figure}[th]
\centerline{\epsfxsize=3.5in\epsfbox{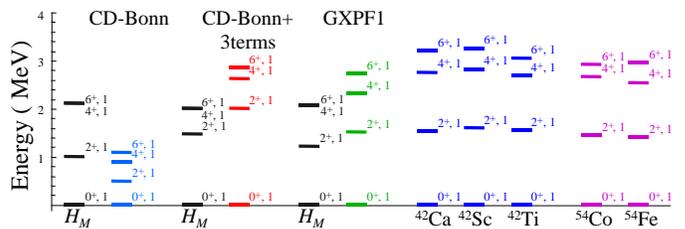}}
\caption{(Color online) Energy spectra of $T=1$ states predicted by 
the \CDB~ (light
blue), \CDBt~
(red) and \Gm~ (green) realistic interactions. Each is compared to the model
Hamiltonian $H_M$ (black) with $\chi =0.071,\ 0.036$ and $0.055$,
respectively (Table \ref{tab:enMoments}). For comparison, the
experimental $T=1$ energy
spectra of the $A=42$  Ca, Sc, Ti isobars (blue) and $A=54$ Co and Fe
isobars (magenta)
are  also shown.}
\label{TscalarSpectra}
\end{figure}

The $T=1$ part of the pure two-body realistic interactions is
reproduced quite well by
the model Hamiltonian. This in turn yields a similar energy  spectra (Figure
\ref{TscalarSpectra}) as predicted by the model interaction $H_M$ and both the
\CDBt~ and \Gm~ realistic interactions. They also agree well with the available
experimental data for the $A=42$ and $A=54$ isobars (the latter
refer to systems of
two holes). Here again, the agreement between $H_M$ and \CDB~ is not
as good as for the
other two interactions, especially in reproducing the spreading  of
the states, which is
smaller for \CDB~ relative to the experimental energy spectra.
Compared to the \Spn{4} model interaction, the energy of the first
$2^+$ state is brought slightly lower by the addition of $\HQ $, to
1.47 MeV (when
$\chi $ is determined in comparison to \CDBt) and 1.22 MeV (to \Gm).
Such values are a bit closer to the experimental results, namely,
1.52-1.59 MeV for the
$A=42$ isobars and 1.41-1.44 MeV for  the $A=54$ isobars in the
$\flevel$ level. While
the influence of the quadrupole-quadrupole  interaction is
significant in the $2^+$ $T=1$
states, it does not affect the estimate for the
\Spn{4} parameters because they were determined with regard to the
nuclear \IASs.

An interesting result is that in the detailed case of isospin-scalar
distribution,
the $(J=0,T=1)$ two-body matrix element is not affected at all by the
$Q \cdot Q$
interaction added to $H_{ \spn{4} }$ because \HQ~  has a zero
contribution to this state.  In this way, the addition of a
quadrupole-quadrupole
interaction does not alter the $0^+$ $T=1$ ground state of the $A=42$ isobars,
which are precisely the lowest \IASs~where the \Spn{4} model has been applied.
Moreover, the average effect of this two-body collective interaction
throughout the
entire shell is zero [see (\ref{HM})]. Even though this property of \HQ~  is
imposed by construction, such a form (as already mentioned) is
typically preferred so as
to  preserve the
shell structure. In short, the domain of states where the
\Spn{4} model was applied is not influenced by the inclusion of the quadrupole
degree of freedom to the pairing model. The \Spn{4} model interaction
itself actually
accounts for all the effects, small or large, due to the influence of the
quadrupole-quadrupole interaction on these states.

\subsection{Correlations between Interactions for Nuclear Systems
with More Than Two Nucleons}

An important feature of spectral distribution theory is that the correlation
coefficient concept can be propagated beyond the defining two-nucleon system to
derivative systems with larger numbers of nucleons \cite{ChangFT71} and in the
isospin-scalar case, for higher values of isospin \cite{HechtDraayer74}. The
propagation formulae (\ref{<HH'>}) determine how the
averages extracted from
the two-nucleon matrix elements in the two-nucleon system get carried
forward into
many-nucleon systems. This propagation of information is
model-independent. In this way one
can track the similarity of pairing/rotational characteristics
between different
interactions in many-nucleon systems \cite{KotaPP80}.

\begin{figure}[th]
\centerline{\epsfxsize=1.0in\epsfbox{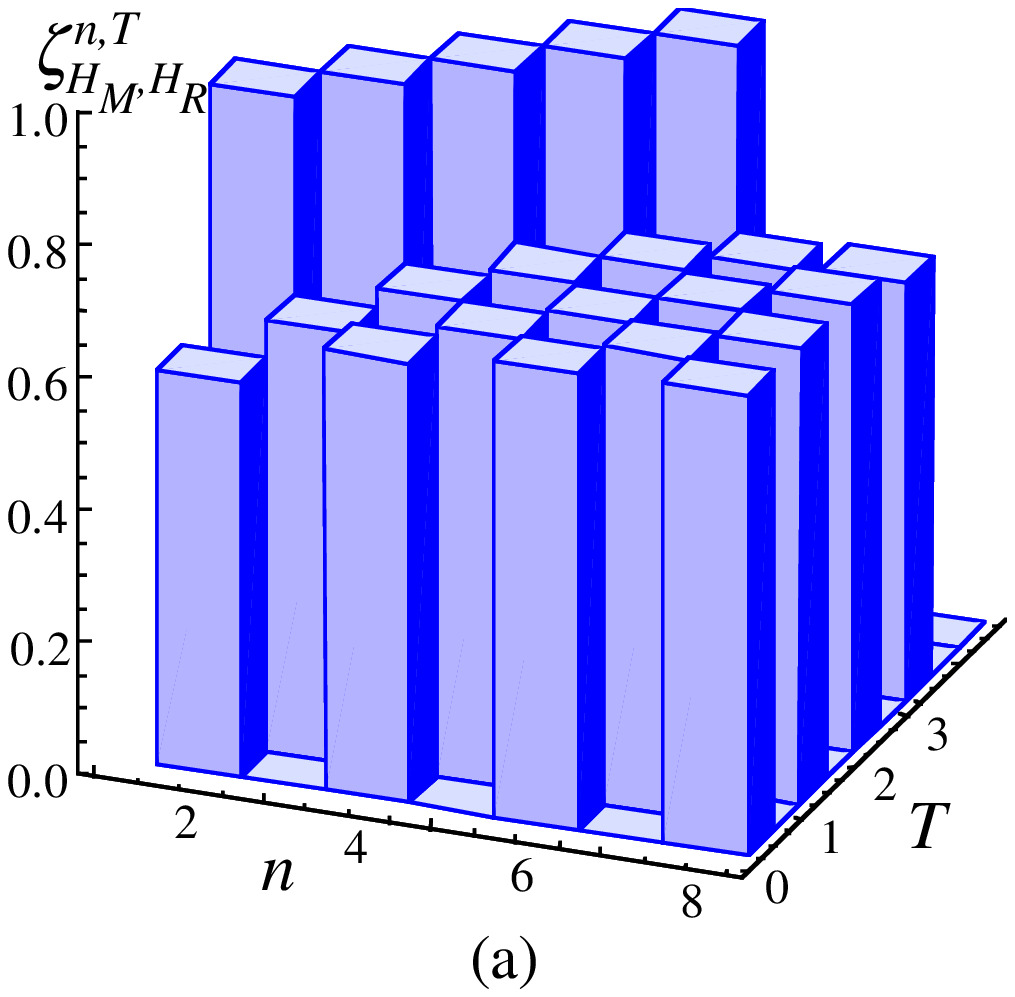}
\epsfxsize=1.0in\epsfbox{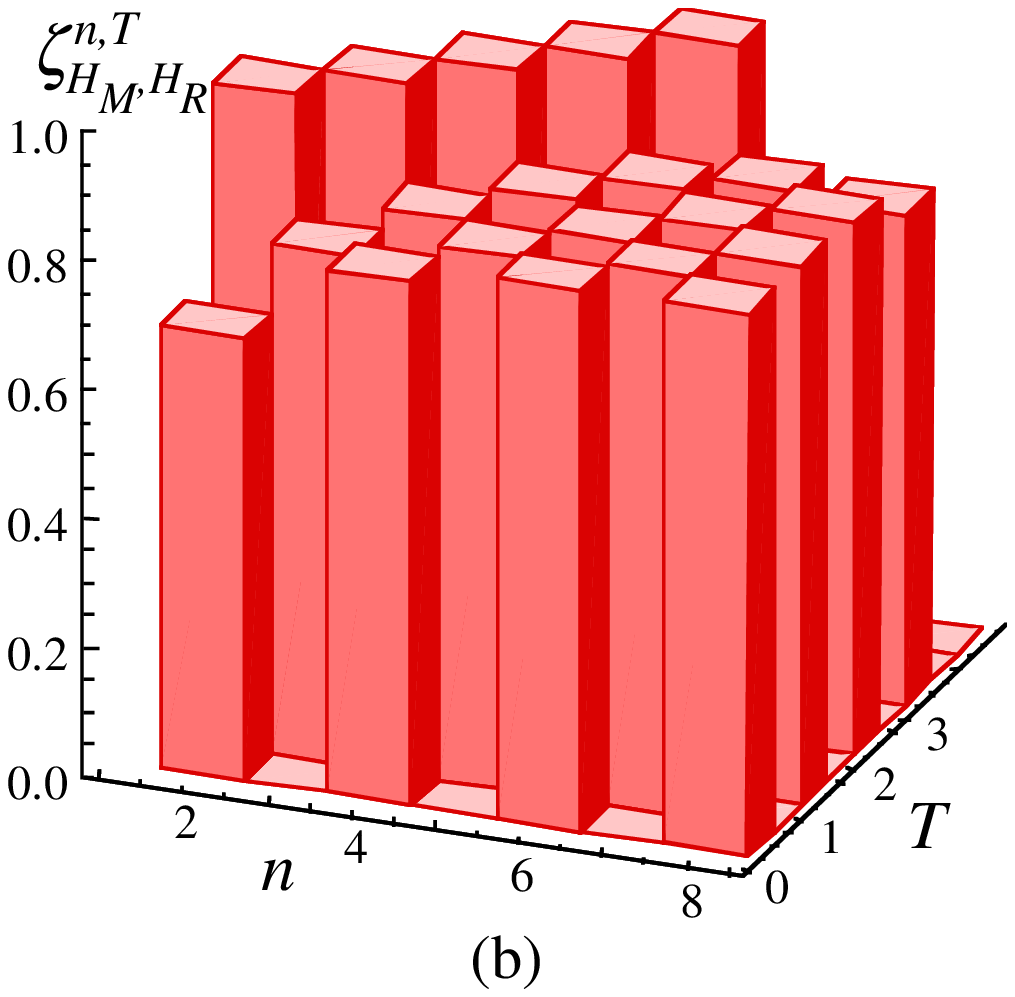}
\epsfxsize=1.0in\epsfbox{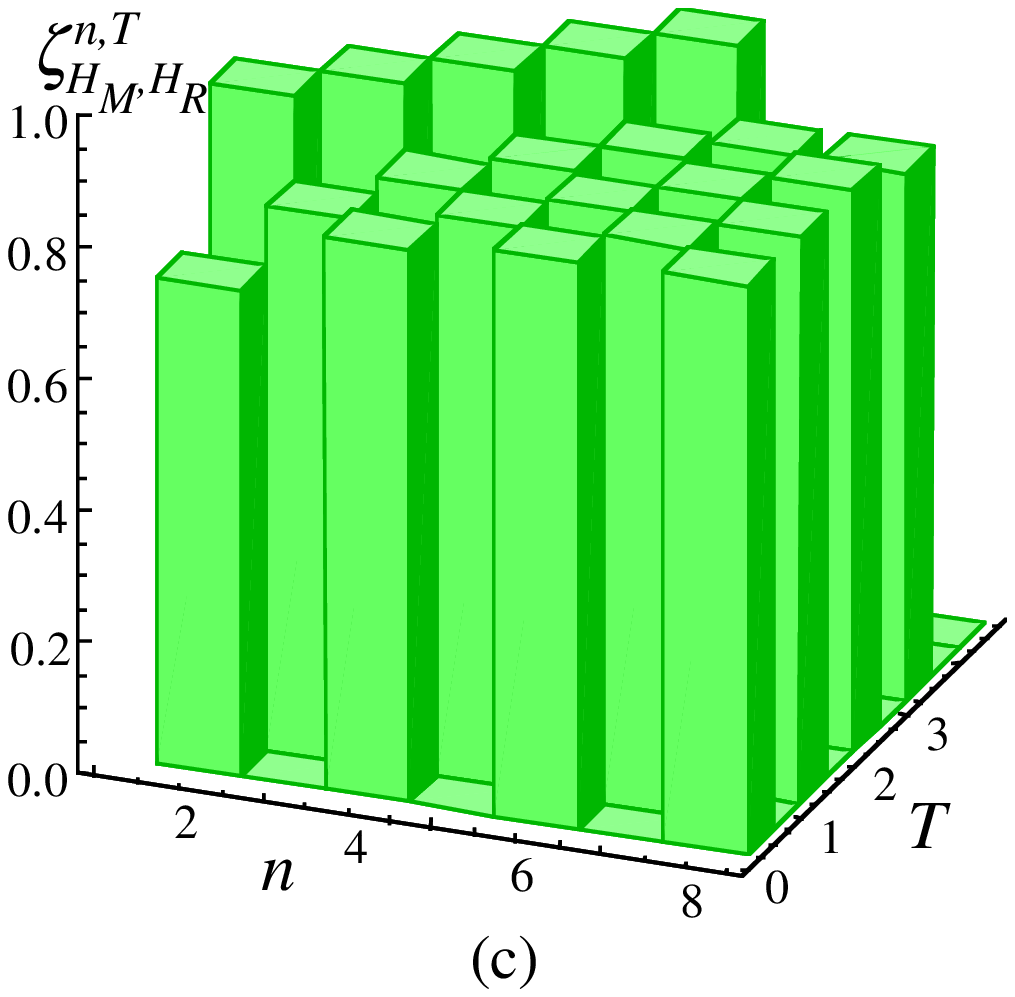}}
\caption{(Color online) Correlation coefficients, $\zeta ^{n,T} 
_{H_{M},H_R}$, of
the pairing+quadrupole
model interaction with (a) \CDB, (b) \CDBt~ and (c) \Gm~ as a
function of $n$ and
$T$ for the \flevel orbit. The representation is symmetric with
respect to the sign
of $n-2\Omega$.}
\label{TscalarCorrelmT}
\end{figure}

In the scalar case the correlations between the interactions retain
their values as given in Table \ref{tab:CC0}.
For the isospin-scalar distribution, the correlation coefficients
between the realistic interactions and the $H_{M}$ model interaction
decrease for
$n>2$ and higher-$T$ values compared to the $n=2$ $T=1$ case and increase when
compared to the $n=2$ $T=0$ estimates (Figure \ref{TscalarCorrelmT}).

For given $n$ and $T$, the $\zeta $ values can
again be found as the maximum correlation for an optimal value of the
quadrupole-quadrupole strength $\chi$, which is related to the
angle between the
geometrically represented $H_R$ and its projection onto the $H_M$ plane.
When $T=n/2$ (Figure \ref{TscalarCorrelmT}),
which corresponds to the highest possible isospin states
(including $n=2,\ T=1$) in 41
nuclei with  valence nucleons occupying
the \flevel orbit, the realistic interactions continue to be strongly
correlated with the
$H_{M}$ model Hamiltonian, namely, $\zeta^{(T)}_{H_{M},H_R}$ is 0.95,
0.98 and 0.96 for \CDB,
\CDBt~ and \Gm, respectively. For these states, the other types of
interactions are  negligible and constitute only 3\% of \CDBt, 8\% of
\Gm~and 9\% of \CDB~ (Figure \ref{Vs}(b)).

For all the cases with $T \ne n/2$ throughout the entire shell, the correlation
coefficients are found to retain almost the same value (Figure
\ref{TscalarCorrelmT}),
namely, $\zeta^{T}_{H_{M},H_R}$ is around $0.63-0.70$, $0.76-0.84$
and $0.80-0.86$ for
\CDB, \CDBt~ and \Gm, respectively, with the corresponding optimal
strength $\chi$ of the quadrupole-quadrupole interaction in the
intervals, $0.087-0.096$, $0.042-0.047$,  and $0.057-0.058$. The 
smaller $\chi$, the
weaker the \Spn{4} symmetry breaking resulting from the additional 
quadrupole-quadrupole
interaction. Again, the least strength is observed  when $H_M$ is 
compared to \CDBt.
In the case of \Gm, $\chi$ remains almost the same for all the 
states, with $T\ne n/2$ as
well as  with  $T=n/2$ (Table \ref{tab:enMoments}, last row). In 
addition, the relative
contribution of other types of interaction, which are not accounted 
for by $H_{M}$
(\ref{HM}), is somewhat greater for \CDBt~ and \CDB~ compared  to 
\Gm~ because of their
comparatively  smaller $\zeta^{T \ne n/2}_{H_{M},H_R}$ (Figure 
\ref{TscalarCorrelmT}). In
short, among the three realistic interactions, when $T$ differs from 
$n/2$ the \Spn{4}
dynamical symmetry continues to be reflected the most in \CDBt, while the
extended pairing+quadrupole model interaction correlates the best with \Gm.

An interesting result is that among the $T \ne n/2$ cases
the highest $\zeta^{T}_{H_{M},H_R}$ correlations for each realistic
interaction are observed for the low-$T$ mid-shell nuclear states, where
the $\chi$ strength is relatively smaller. This suggests that for these
states the other kinds of interactions not present in $H_M$ grow weaker and the
realistic interactions are comparatively closer in behavior to the
symplectic pairing \Hsp. The lowest correlation is observed for the
mid-shell nuclei
with isospin next to the highest, where $\chi $ is the largest. For
these cases, both
quadrupole-quadrupole and other types of interactions accounted for
by  the realistic
interactions increase in relative importance.

In summary, for all of the $(n,T)$ distributions the model
pairing$+$quadrupole
interaction accounts on average for about 59\%, 77\% and 78\% of the
\CDB, \CDBt~ and
\Gm~realistic interactions, respectively, and up to 91\%,  97\%, and
92\% of those in
the highest possible isospin states for a given $n$, where the $H_{M}$ model
interaction can be  used to provide a reliable description.
\newline

Although, the aforementioned results refer to a single-$j$ orbit ($\flevel$),
they represent a first step towards a generalization to multi-$j$ major shells
(such as the $fp$ shell or even a set of several major shells). Such
an extension is certainly feasible and not space limited from the
perspective of spectral
distribution theory. Even though the statistics for the
\flevel level are not large compared to the $fp$ shell, it is a
natural choice. Specifically, the \flevel orbital is comparatively far away
from both the
neighboring $ds$ and upper $fp$ shells which means there is a
preponderance of a
single-$j$ coherence over configuration mixing in the low-lying
nuclear states for
such nuclei. For this reason it is an interesting example in its simplicity and
provides a quite clear view of the pairing/rotational foundation of the nuclear
interaction that is free of competing configuration mixing effects.
In addition, it does represents a partitioning of the $fp$-space and
as such it provides
for more detailed spectral measures that may reflect important fine
effects that are
otherwise averaged out when the entire $fp$ major shell is taken into account.
It will be also important to augment this single-$j$ shell
study with a
follow-on, complementary multi-shell analysis of similar type.  The
multi-shell case will be the topic of a future publication.

In addition to the above, we note for completeness that the estimates
for the parameters
of the model \Spn{4} interaction are not really  relevant to the
primary objective of the
present study. The reason is that the correlation  coefficients in the detailed
isospin-scalar case are independent of the interaction strength
parameters, which
therefore do not affect the  correlation measures of the model
interaction with realistic
ones or the pairing and rotational characteristics of the latter.

\section{Conclusions}
With a view towards a broader study within multi-$j$ shells, we
compared three realistic
interactions and two model pairing and quadrupole  interactions for the
\flevel orbit by means of the theory of spectral distributions.

In the more detailed case of isospin-scalar distribution, the \CDB,
\CDBt~ and \Gm~ realistic interactions were found to  contain
on average 59\%, 77\%, and 78\%, respectively, of the 
pairing$+$quadrupole interaction.
Moreover, this
percentage  goes up to 91\%, 97\%, and 92\%, respectively, for the
highest possible isospin
group of states for all the nuclei  with valence protons and neutrons
occupying the \flevel
shell. For these states, the strongest  correlation was observed
between the \CDBt~ and the
pairing$+$quadrupole model interaction, where other types of
interaction accounted in the
realistic  interactions represent only 3\% of it. They constitute 8\%
of the \Gm~ realistic
interaction, and 9\% of \CDB.
For these cases, the pairing$+$quadrupole model
interaction has been shown to be a very good approximation that provides a
reasonable description of the energy spectra of the nuclei in the
\flevel level. While both
interactions,
\CDBt~ and \Gm, exhibit a well-developed pairing character compared
to \CDB, the latter
appears to build up more (less) rotational collective features that
are outside of the
scope of the $T=1$ ($T=0$) \Spn{4} interaction.

The major advantage of the \spn{4} algebraic model, which focuses on
the isovector
pairing correlations and also includes a certain portion of the
quadrupole-quadrupole
interaction, is that it provides an elegant solution for describing
the pairing-governed
\IASs~ in light and medium mass nuclei. In addition, while it
correlates to a reasonably
good extent with the realistic interactions, the description of the
low-lying nuclear
energy spectrum of higher-$J$  states is improved with the inclusion of the
quadrupole-quadrupole interaction that  being symmetric under
\SU{}{3} breaks the
\Spn{4} symmetry and removes degeneracies.  Nevertheless, we found that for the
isospin-scalar distribution the \Spn{4} model  interaction accounts
for a large part of
the \CDBt~ realistic interaction. It also  includes between 15\% to 35\% of the
rotational collective interaction and typically  accounts for a
rather large portion of
the overall correlation of the realistic interactions with the
pairing$+$quadrupole
interaction.  Moreover, the additional  quadrupole degree of freedom
incorporated in the
symplectic model interaction does not  affect the domain of states
where the \Spn{4}
model was applied and hence introduces no  errors in the estimates of
the parameters of
the symplectic interaction. These  results confirm the conclusion
that the \Spn{4}
interaction can provide for an  approximate pattern of the nuclear
energy spectra and,
above all, can be accepted as a very reasonable approximation to describe the
pairing-governed \IASs~ in the nuclei in the \flevel orbit.

\section*{Acknowledgments}

%\vskip .5cm
This work was supported by the US National Science Foundation, Grant
Number 0140300
and the Southeastern Universities Research Association.

\end{document}